\documentclass{jnmauth}
\usepackage{amsmath}
\usepackage{amssymb}
\usepackage{graphicx}
\usepackage{bifonts}
\usepackage{color}

\newcommand{\mat}[1]{\boldsymbol{\uppercase{#1}}}

\renewcommand{\mat}[1]{\uppercase{\mathbi{#1}}}             
\newcommand{\mleft}[1]{\left[\!\begin{array}{#1}}
\newcommand{\mright}{\end{array}\!\right]}

\renewcommand{\j}{\mathrm{j}}
\newcommand{\e}{\mathrm{e}}

\begin{document}
\JNM{1}{6}{00}{28}{00}

\runningheads{K.\ Ochs et al.}{Anticipation of digital patterns}

\title{Anticipation of digital patterns}

\author{Karlheinz Ochs\,$^{1,*}$, Martin Ziegler\,$^{2}$, Eloy Hernandez-Guevara\,$^{1}$, Enver Solan\,$^{1}$,  Marina Ignatov\,$^{2}$, Mirko Hansen\,$^{2}$, Mahal Singh Gill\,$^{2}$, and Hermann Kohlstedt\,$^2$}
\address{$^{1}$Chair of digital communication systems, Electrical engineering and information technology, Ruhr-Universit\"{a}t Bochum, 44780 Bochum, Germany\\
	     $^{2}$Nanoelektronik, Technische Fakult\"{a}t, Christian-Albrechts-Universit\"{a}t zu Kiel, 24143 Kiel, Germany}

\corraddr{Karlheinz Ochs, Department of Electrical Engineering and Information Technology/Communications Engineering,
Ruhr-University, D-44780 Bochum, Germany; ochs@ieee.org}

\received{14 10 2016}
\revised{dd mm yyyy}
\noaccepted{}

\begin{abstract}
A memristive device is a novel passive device, which is essentially a resistor with memory. This device can be utilized for novel technical applications like neuromorphic computation. In this paper, we focus on anticipation -- a capability of a system to decide how to react in an environment by predicting future states. Especially, we have designed an elementary memristive circuit for the anticipation of digital patterns, where this circuit is based on the capability of an amoeba to anticipate periodically occurring unipolar pulses. The resulting circuit has been verified by digital simulations and has been realized in hardware as well. For the practical realization, we have used an Ag-doped TiO$_{2-x}$-based memristive device, which has been fabricated in planar capacitor structures on a silicon wafer. The functionality of the circuit is shown by simulations and measurements. Finally, the anticipation of information is demonstrated by using images, where the robustness of this anticipatory circuit against noise and faulty intermediate information is visualized.
\end{abstract}


\section{INTRODUCTION}
Even unicellular organisms like amoebas exhibit astonishing behavioral capabilities. For example, initially motivated by foraging, the amoeba Physarum polycephalum is able to solve mazes and in particular geometrical puzzles~\cite{NYT00, NYH04}. In an experiment, Saigusa et al. demonstrated anticipation in amoeboid organisms, which were exposed to periodic environmental events~\cite{STNK08}. Herein, the organism learned a periodic shock event pattern and anticipated a further subsequent event, although the event was experimentally not applied. Anticipation is not restricted to cellular organisms and is understood as a very general concept in nature \cite{Ros12}. The various aspects of anticipation have been debated in numerous research fields as for example, natural sciences, social sciences, and engineering~\cite{Ros12, Pol10}. A well-accepted definition of the term "{anticipation}" was given by Rosen and reads:\textit{ An anticipatory system is a system containing a predictive model of itself and/or of its environment, which allows it to change at an instant in accord with the model's predictions pertaining to a later instant}. Anticipation is not restricted to living species. Under the framework of bio-inspired engineering, anticipation is considered as an interesting approach, e.g.  for cell-based robot controller or multi-agent robotics~\cite{TZG07, Tan07}.
The above-mentioned amoeba experiment by~\cite{STNK08} was later modeled by~\cite{PLD09} based on an electric resonance circuit comprising a memristive device. The model exhibited electronically essential features of the biological experiment. In 2012 anticipation, as inspired by the amoeba experiment, was realized experimentally~\cite{ZOHK13}. In this work we present a compact memristive circuit for the anticipation of digital patterns, which extends the amoeba learning circuit \cite{PLD09} to an anticipatory system for bipolar signals. After a quantization procedure, this anticipatory system can be used for the anticipation of information represented by general bit patterns. This allows for a variety of possible applications, especially from the field of bio-inspired engineering.

\section{The anticipation system}

\subsection{Anticipation circuit for analogue patterns}
\label{ssc:Anticipation circuit for analogue patterns}

\begin{figure}[!hbt]
	\centering
	\includegraphics[scale=0.75]{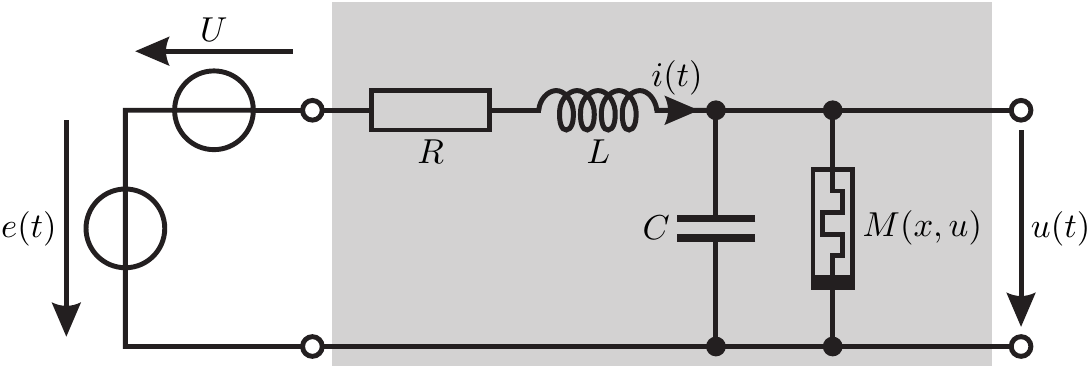}%
	\caption{Anticipation two-port driven by two voltage sources, cf.~\cite{ZOHK13}.}
	\label{fig:AnticipationCircuit}
\end{figure}

Anticipation events, as it has been observed in unicellular organisms like amoebas~\cite{STNK08}, was mimicked by a memristive resonance circuit~\cite{PLD09} as shown in Fig.\,\ref{fig:AnticipationCircuit}.

The circuit comprises a series interconnection of two voltage sources, which supply an anticipation two-port (grey-shaded) with the signal $e(t)+U$. Here, $e(t)$ is an information-bearing input voltage and $U$ is a positive voltage offset. The anticipation two-port is a damped oscillator, including a resistance, a capacitance, an inductance, and comprises especially a memristive device \cite{Chua71, Chua76}:
\begin{align}
   u &= M(x,u)\,i\:, \qquad
   \dfrac{dx}{d t} = f(x,u)\:.
\label{memdef}
\end{align}
The structural parameter $x$ represents the memory behavior of the memristive device, and is itself a suited function of the applied voltage $u$. Independent of the specific function~$f$ the memristive device is due to its non-negative memristance~$M$ a passive one-port \cite{Chua71}. Due to the fact that only passive one-ports are port-wise interconnected via a Kirchhoff network, the anticipation two-port is itself passive.

\pagebreak

Most of the memristive devices are based on ionic transport mechanisms \cite{RRS09} and are consisting of the layer sequence metal-insulator-metal. The memristive behavior is based on an ionic drift inside the insulator layer under the applied electrical field. In those devices, $f$ is physically related to an ionic drift model~\cite{BBB09}. To be more specific, here we have used an Ag/TiO$_{2-x}$/Al memristive cell, where the device resistance varies according to the Ag drift/diffusion process. This switches the device resistance at the positive threshold voltage~$U_\mathrm{on}$ from a high ohmic state~$M_\mathrm{off}$ to a low ohmic state~$M_\mathrm{on}$ and vice versa at the negative threshold voltage~$U_\mathrm{off}$, i.\,e. the memristance varies between the low and high ohmic state,
\begin{align}
   M_\mathrm{on} \le M(x,u) \le M_\mathrm{off}\:,
\end{align}
where $x$ changes accordingly to the differential equation in (\ref{memdef}). Below the threshold voltage $U_\mathrm{on}$ and above the threshold voltage $U_\mathrm{off}$ the memristive device behaves as a constant resistor.

For an electronic implementation of anticipation the circuit must process specific (periodic) input patterns by precisely adjusting the memristive device resistance and simultaneously leaving the resonant frequency unaffected~\cite{ZOHK13}. This is an implicit task because the memristive device, in turn, must be adapted to the circuit oscillations evoked by the input pattern. An analytical description of the memristive anticipation two-port is hindered by the recursive dependency of the memristance $M(x,u)$ from the state variable $x$ and the voltage $u$, cf.~\cite{ZOHK13}\:. Nevertheless, under the restriction of a constant memristance~$M$, it is possible to get a general insight into the anticipation two-port functionality. Whereas in~\cite{ZOHK13} one can find a steady-state analysis of the anticipation circuit, we will here investigate the transient behavior by analyzing the pulse response of the system, which is important for the anticipatory functionality of the system. To this end, we start with the system matrix and transfer function of the circuit,
\begin{align}
   \mat{A}
   	&=
   	\begin{bmatrix}
   	-R/L & -1/L\\
   	1/C  & -1/MC
   	\end{bmatrix} &
   & \text{and} &
	H(p)
	&= \frac{\omega_0^2}{[p-\lambda][p-\lambda^*]}\:,
\end{align}
respectively \cite{ZOHK13}. Here, $p$ is the complex frequency variable, $\omega_0=\frac{1}{\sqrt{LC}}$ is the resonance frequency of the ideal $LC$-circuit. The eigenvalues of this real damped oscillatory system are complex conjugate pairs $\lambda_1 = \lambda_2^*=\lambda=-\gamma+\j\omega_\mathrm{r}$\:, where
\begin{align}
	\gamma
	&= \frac{\omega_0}{2}\left[r+\frac{1}{m}\right]\:,\quad\omega_r^2 = \omega_0^2\left[\frac{r}{m}+1\right]-\gamma^2\:,\:\text{with}\:\: r=\frac{R}{Z}\:,\:\: m=\frac{M}{Z} \:\:\text{and}\:\: Z=\sqrt{\frac{L}{C}}\:.
	\label{eqn:complexEigenvalue}
\end{align}
As can be taken from the real and imaginary part, the memristance $M$ affects the damping and the resonance frequency. Especially, these effects are minimal, when the memristive device is in a high ohmic state since $r/m=R/M$ is much smaller than one. In this case the anticipation two-port is able to anticipate pulse trains matched to the resonance frequency
\begin{align}
	e(t)
	&= -e_0 \sum_\mu q(t-t_\mu)\:,
	\quad \text{with} \quad
	q(t)
	= \left\{\begin{array}{rcl}
	1 & \text{for} & 0 \le t \le \frac{T_\mathrm{r}}{2}\\
	0 & \text{otherwise}
	\end{array}\right.\:.
   \label{eqn:pulsetrain}
\end{align}
For a fixed memristance the response of the circuit to this pulse train reads:
\begin{align}
	u(t)
   	&= -e_0
         \sum_\mu \operatorname{Re}\left\{
            \frac{2\,\omega_0^2}{\left[\lambda+R/L\right]^2-\omega_0^2}\,
            \frac{\lambda+R/L}{\lambda}
            \left[1-\e^{-\lambda\frac{T_r}{2}}\right]
            \e^{\lambda\left[t-t_\mu\right]}
         \right\}\:.
\end{align}
A detailed examination of this response shows that the pulses must have negative amplitude~$-e_0$ and ideally they should equally spaced by the resonant period $T_r$, with $\omega_\mathrm{r} T_\mathrm{r} = 2\pi$. Pulses at these instants lead to a constructive oscillating behavior where the output voltage~$u(t)$ has a high amplitude, which in turn indicates anticipation.

\subsection{Anticipation circuit for digital patterns}
The anticipation two-port is only able to anticipate pulses with negative amplitude, i.\,e. the anticipation is limited to unipolar signals. The aim of this section is to extend the circuit to bipolarity. For this purpose, the bits $0$ and $1$ are mapped to $-e_0$ and  $+e_0$, respectively:
\begin{align}
	e(t)
	&= e_0 \sum_{\mu} x_\mu\,q(t-t_\mu)\:,
	\quad \text{with} \quad
	x_\mu
	= \left\{\begin{array}{rcl}
	-1 & \text{for} & b_\mu=0\\
	+1 & \text{for} & b_\mu=1
	\end{array}\right.\:,
	\label{eqn:bittrain}
\end{align}
where $b_\mu$ denotes the bit associated with the instance~$t_\mu$. In particular, a digital anticipator should allow for anticipating the bits by generating a tristate output signal, which indicates whether bit $1$, bit $0$ or nothing has been anticipated.

In order to anticipate positive pulse trains, one can modify the electrical circuit of Fig.~\ref{fig:AnticipationCircuit} by interconnecting an ideal transformer with turns ratio $1:-1$ between the information bearing input voltage source~$e(t)$ and the rest of the circuit. Because of the sign inversion with respect to the voltages of the primary and secondary side of the transformer, this modification leads to an anticipator of positive pulses. Next, we can get rid of the ideal transformer by transforming the complete circuit on the secondary side. This solely results in a sign inversion of the constant voltage source~$U$ and an interchanged orientation of the memristive device.

\begin{figure*}[!hbt]
	    \centering
		\includegraphics[width=1\linewidth]{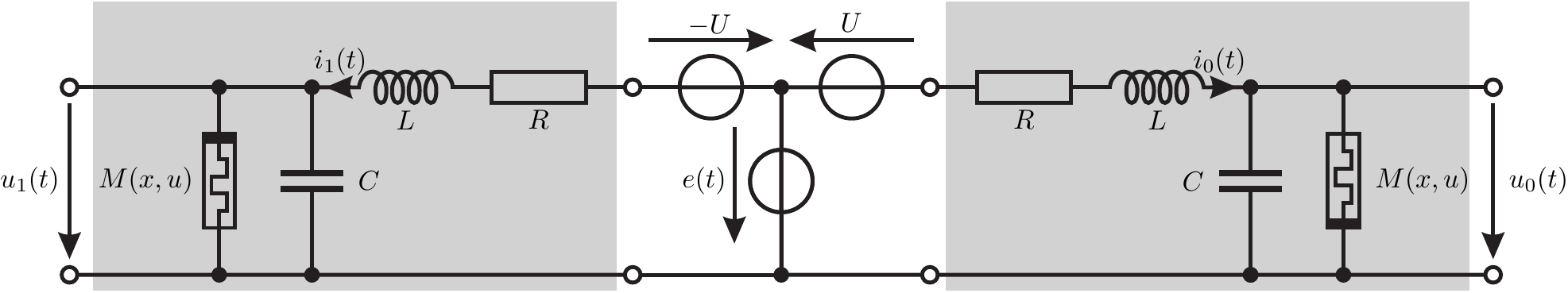}%
		\caption{Bit anticipator circuit.}
		\label{fig:BitAnticipator}
\end{figure*}

A combination of both anticipation circuits leads to an electrical circuit layout as sketched in Fig.~\ref{fig:BitAnticipator}, where the left and right part anticipates positive and negative pulse trains, respectively. For this circuit, a negative pulse train will be anticipated by the right anticipator and yields an oscillating voltage $u_0(t)$. In contrast to this, a positive pulse train will be anticipated by the left anticipator, which results in an oscillating voltage $u_1(t)$. Hence, the circuit layout depicted in Fig.~\ref{fig:BitAnticipator} is able to anticipate bipolar signals and will be called bit anticipator in the sequel.

\begin{figure}[!hbt]
     	\centering
		\includegraphics[scale=0.9]{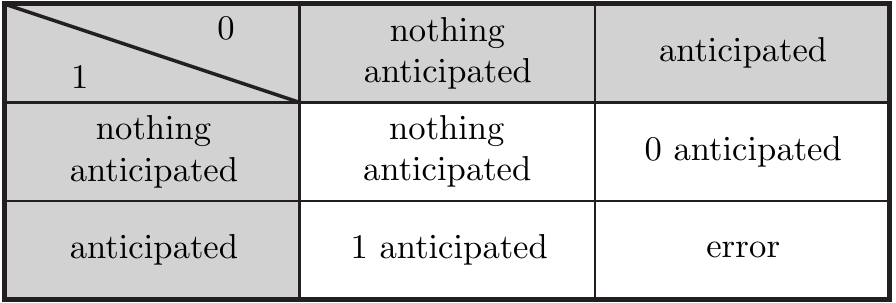}%
		\caption{Logical states for the anticipation of digital pattern.}
		\label{fig:BitAnticipationTable}
\end{figure}

It remains to define the different logical states of the digital anticipator, which are tabulated in Fig.~\ref{fig:BitAnticipationTable}: The bit anticipator anticipates nothing if neither the left nor the right anticipator anticipates. It anticipates a $0$ respectively a $1$ if only the associated output voltage is oscillating. The situation is critical if both anticipators have an oscillating output signal because anticipating a bit~$1$ and a bit~$0$ simultaneously is obviously an error. This error can be treated in many ways. For instance, we can decide that the circuit ignores this case or both anticipators are reset to their initial states.

\section{Hardware design}
In the following section, the technical realization of the bit anticipator is presented. In Sec.~\ref{sec:MemristiveDevices} fabrication issues and the resistive switching behavior of the used memristive devices are discussed, while Sec.~\ref{sec:CircuitImplementation} deals with the hardware design of the bit anticipator.

\subsection{Memristive Device}\label{sec:MemristiveDevices}%
The key components of the anticipation circuit are memristive devices, which we have externally connected to the circuit board. The used memristive devices consist of the layer sequence Ag/TiO$_{2-x}$/Al and are fabricated as planar capacitor structures on a Si-wafer with $400\,\mathrm{nm}$ of thermally oxidized SiO$_2$ by employing a standard optical lithography process. In detail, Ag is deposited by thermal evaporation. In a one-step lithography $50\,\mathrm{\mu m} \times 50\,\mathrm{\mu m}$ contact windows are defined, followed by reactive ion sputtering of $10\,\mathrm{nm}$ TiO$_{2-x}$ and $5\,\mathrm{nm}$ Al. For the wiring $140\,\mathrm{nm}$ Al is deposited by thermal evaporation on top, while the device fabrication is completed by a subsequent lift-off in acetone.

The fabricated devices belong to the class of electrochemical metallization cells (ECM) for which the change in resistance is due to the formation and dissolution of electrically conducting paths in a transition metal oxide~\cite{RRS09,SMM15}. On a macroscopic scale, the mechanism of resistive switching is typically illustrated as follows: Metal atoms from the active electrode are oxidized and drift/diffuse through the transition oxide layer towards the opposite electrode.

\begin{figure*}[!hbt]
	\begin{center}
		\includegraphics[width=0.45\textwidth]{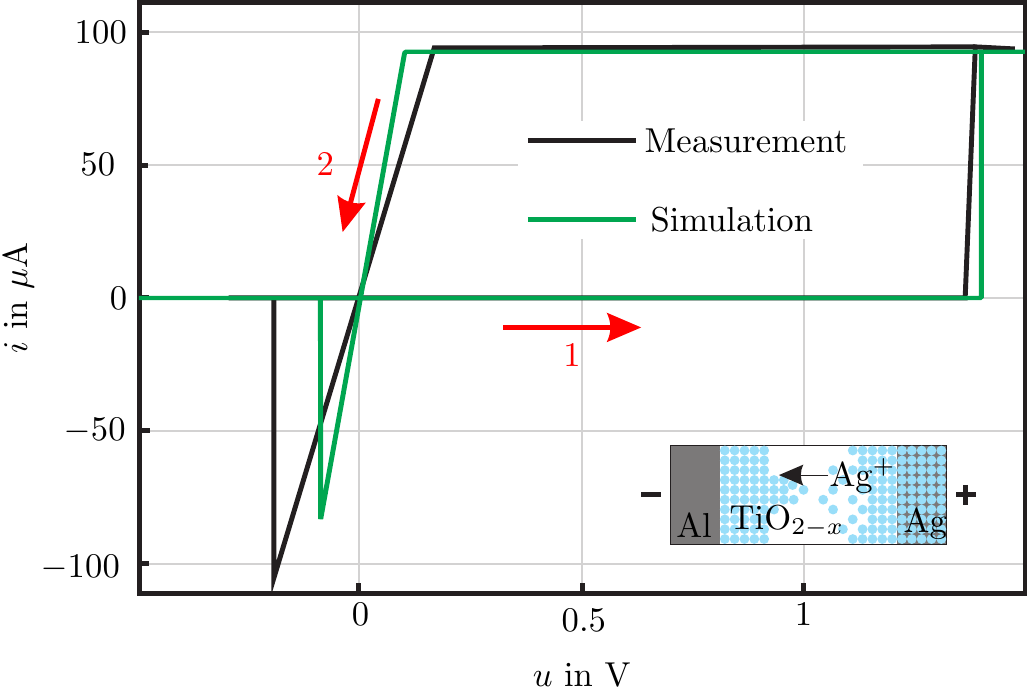}\qquad%
		\includegraphics[width=0.45\textwidth]{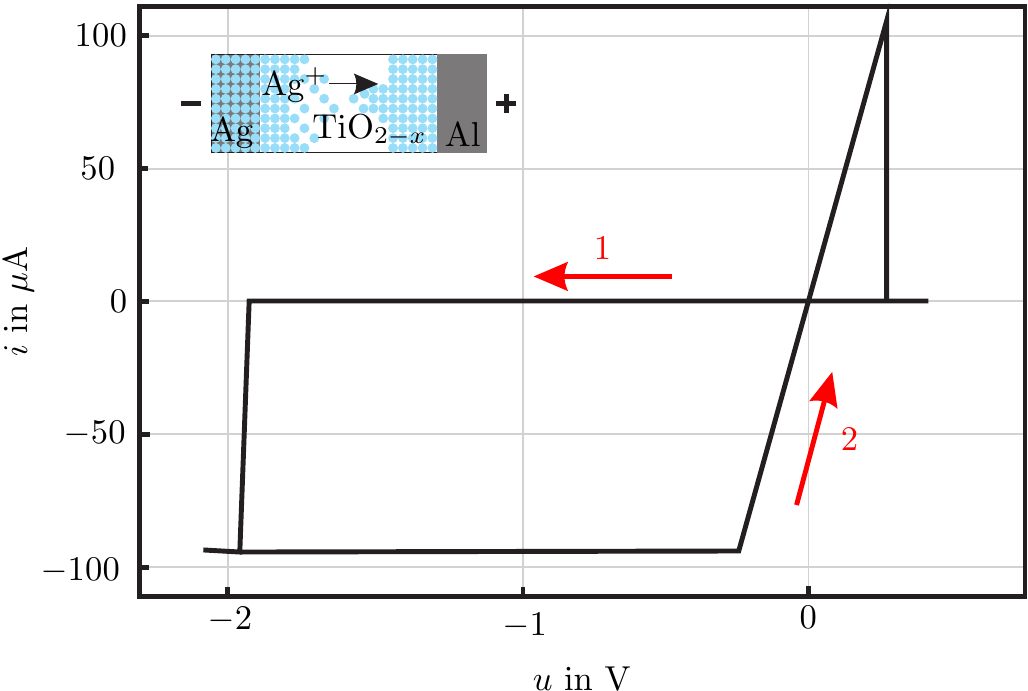}%
	\end{center}
	\caption{Typical current-voltage characteristics (i-u-curves) of Ag-doped TiO$_{\text{2-x}}$-based memristive cells for different electrical connections. While in (left) negative and positive voltages are, respectively, applied through the left and right electrode, in (right) the device polarity is reversed. For the measurement a compliance current of 100 $\mu$A has been used. Insets: Schematic of the layer sequence of the Al/TiO$_{\text{2-x}}$/Ag device indicating the electrical connection. For a comparison, the simulated curve of the memristive model used in simulations is also shown in the left figure (green).}
	\label{fig:ExperimentHysteresis}
\end{figure*}

Here, they can be reduced and start clustering. Step by step conductive filaments are growing and binary bipolar resistive switching characteristics are obtained. In Fig.~\ref{fig:ExperimentHysteresis} two typical current-voltage characteristics (i-u-curves) for two individual Ag/TiO$_{2-x}$/Al devices are shown where the connections to the external voltage source are interchanged in order to reach the requirements defined by (\ref{eqn:bittrain}). For this measurements, an Agilent E5260 source measurement unit was employed by sweeping the applied voltage and measuring the device current, simultaneously. Moreover, a current compliance of $100\,\mathrm{\mu A}$ has been set in order to avoid an electrical break down of the memristive cell. In particular, resistive switching from the initial high resistance state~$M_\mathrm{off}$ to the low resistance state~$M_\mathrm{on}$ is observed if a positive voltage is applied to the active Ag-electrode, cf. Fig.~\ref{fig:ExperimentHysteresis}. For the back transition from the $M_\mathrm{on}$-state to the $M_\mathrm{off}$-state a negative voltage is required at the Ag-electrode. While $M_\mathrm{off}$ of the fabricated devices is in the range of several $\mathrm{G \Omega}$, $M_\mathrm{on}$ is approximately $10\,\mathrm{k \Omega}$. However, we would like to mention that the $M_\mathrm{on}$ can be varied by appropriate settings of the employed current compliance.

\pagebreak

In general, an important characteristic of ECM cells is that they feature threshold voltages $U_\mathrm{on}$ and $U_\mathrm{off}$ for the resistance set and reset process, respectively. Depending on the voltage source polarity the investigated memristive cells in Fig.~\ref{fig:ExperimentHysteresis} have different threshold voltages: $U_\mathrm{on}=1.4\,\mathrm{V}$, $U_\mathrm{off}=-0.2\,\mathrm{V}$ (left) and $U_\mathrm{on}=-1.9\,\mathrm{V}$, $U_\mathrm{off}=0.2\,\mathrm{V}$ (right). It is worth mentioning that the used class of memristive devices feature an inherent stochastic nature in which the threshold voltages reflect the local atomistic Ag drift/diffusion processes~\cite{SMM15,GSS13}. Thus, such a device stochastically influences the transient dynamic of the bit anticipator circuit. This has to be taken into account for a technical realization.

\subsection{Circuit Implementation}\label{sec:CircuitImplementation}%
\begin{figure*}[!hbt]
    	\centering
		\includegraphics[width=0.6\linewidth]{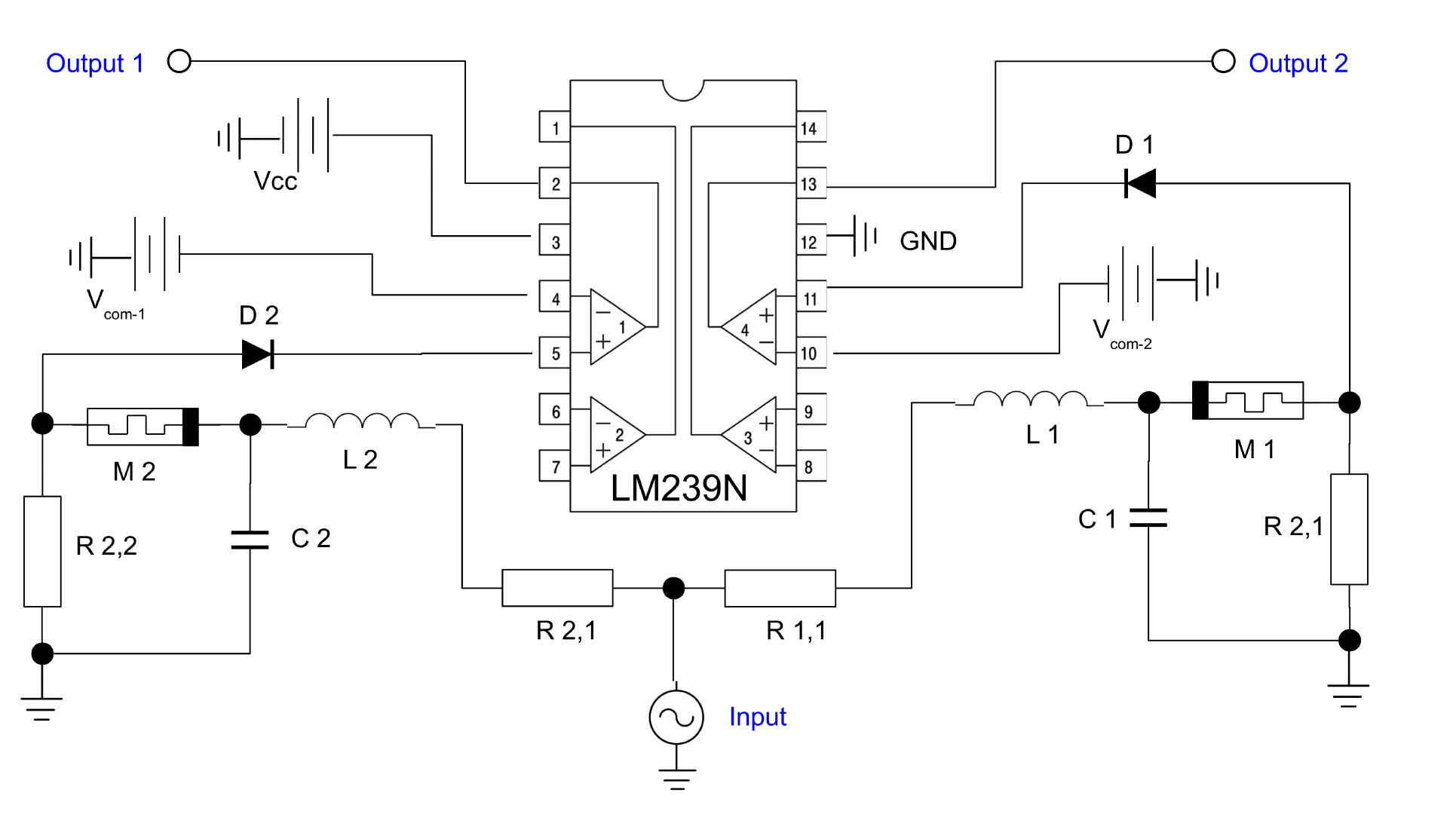}%
		\caption{Circuit layout of the bit anticipation circuit of Fig.~\ref{fig:BitAnticipator}. For the implementation on a printed circuit board the following values of the passive circuit elements have been used: $R_{1,1}=R_{1,2}=100\,\Omega$, $L=100\,\mathrm{mH}$, $C= 150\,\mathrm{nF}$, $R_{2,1}=R_{2,2}=10\,\mathrm{k\Omega}$, and $\omega_0=8.16\,\mathrm{rad/s}$. In addition, a comparator electronic (LM239N) digitizes the anticipated logical states. The comparator threshold voltages are set to $V_{\mathrm{com}}=V_{\mathrm{com-1}}= V_{\mathrm{com-2}}=-100\,\mathrm{mV}$.}
		\label{fig:ExperimentCircuit}
\end{figure*}

The bit anticipation circuit of Fig.~\ref{fig:BitAnticipator} was technical realized on a printed circuit board using the memristive cells of Fig.~\ref{fig:ExperimentHysteresis} and a comparator-electronic to digitize the anticipated output signals $u_0$ and $u_1$, see Fig.~\ref{fig:ExperimentCircuit}. Due to technical reasons, there are slight differences between the realized and idealized circuit mentioned before. The required values for the used passive circuit elements has been taken from~\cite{ZOHK13}.

\section{Results}

\subsection{Simulations}
For a verification of the functionality of this anticipation circuit, we have used LTSpice. The fast switching behavior of real ECM cells has been incorporated by the voltage-controlled memristor model from~\cite{PLD09}. Based on the insights of~\cite{PLD09}, we have adapted the memristor model to be closer to the real device functionality, as depicted in Fig.~\ref{fig:ExperimentHysteresis} (left). Therefore, we have utilized asymmetric thresholds for positive $U_\mathrm{on}$ and negative $U_\mathrm{off}$ applied voltages with
\begin{align}
\begin{aligned}
	f(M,u)
	&= g(u)\,[\sigma(u)\,\sigma(M-M_\mathrm{on})+\sigma(-u)\,\sigma(M_\mathrm{off}-M)]\:, \\
	\text{with}\quad
	g(u)
	&= \beta\left[\sigma(u-U_\mathrm{on})\left[U_\mathrm{on}-u\right]+\sigma(U_\mathrm{off}-u)\left[U_\mathrm{off}-u\right]\right]\:,
\end{aligned}
\end{align}
where $\sigma(u)$ is the unit-step function.

\begin{figure*}[!hbt]
	\begin{center}
		\includegraphics[width=0.4\textwidth]{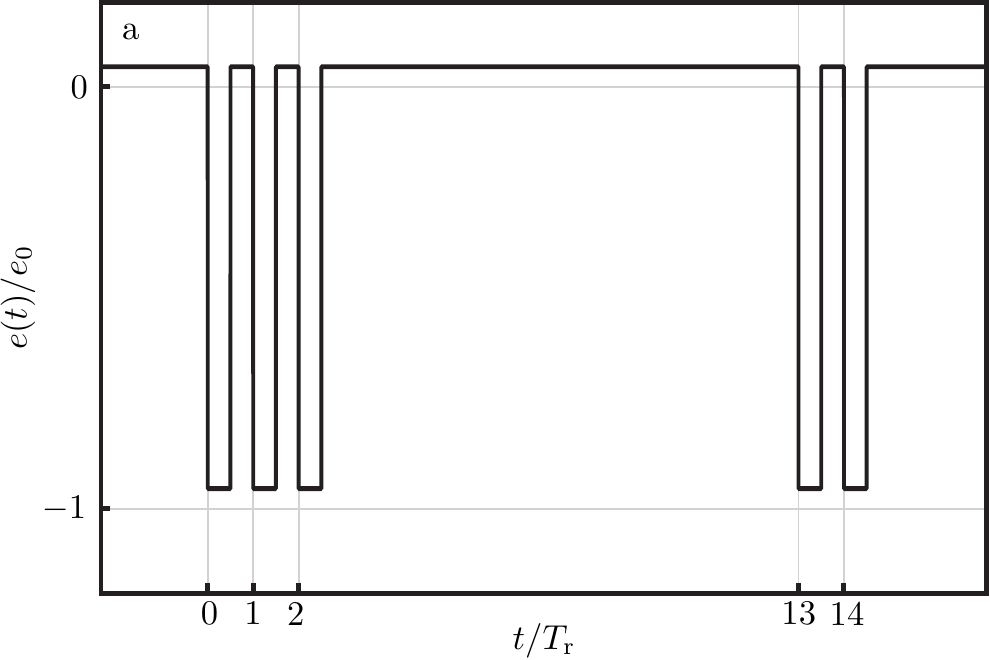}\qquad%
		\includegraphics[width=0.4\textwidth]{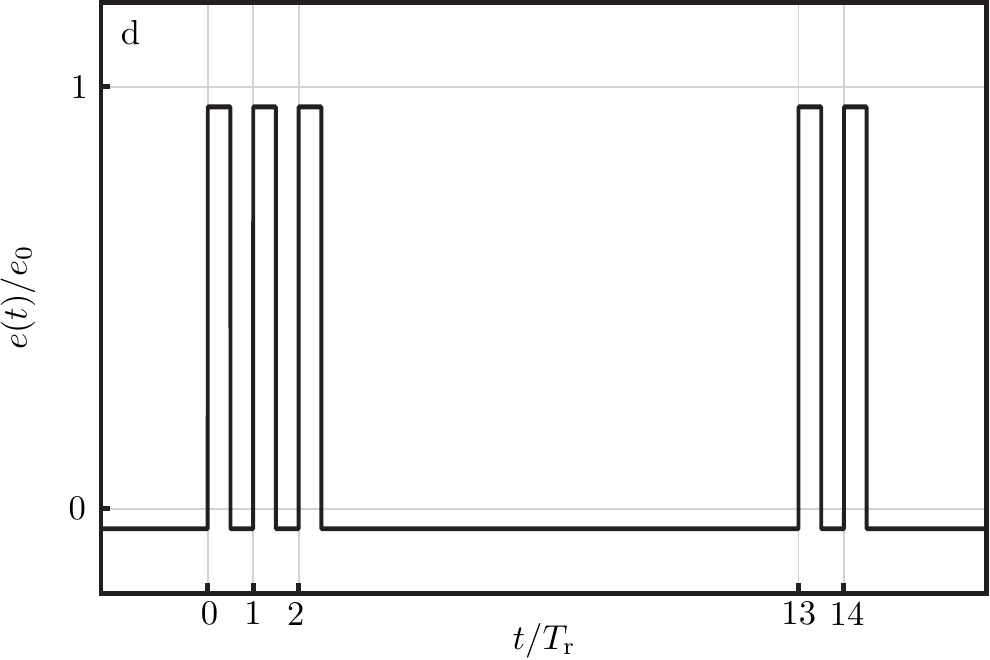}\\[0.45cm]%
		\includegraphics[width=0.4\textwidth]{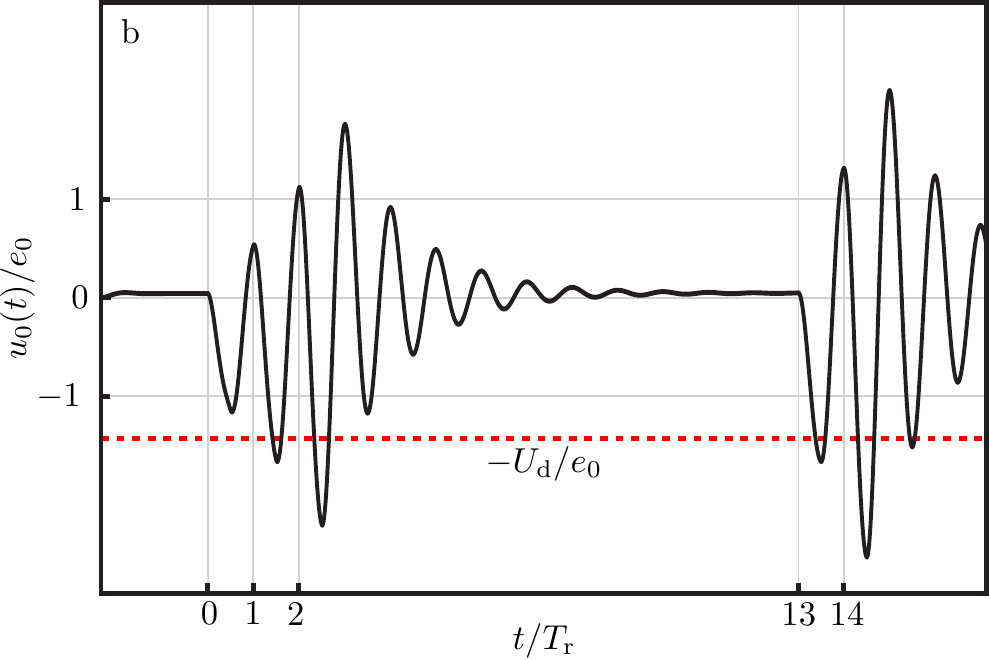}\qquad%
		\includegraphics[width=0.4\textwidth]{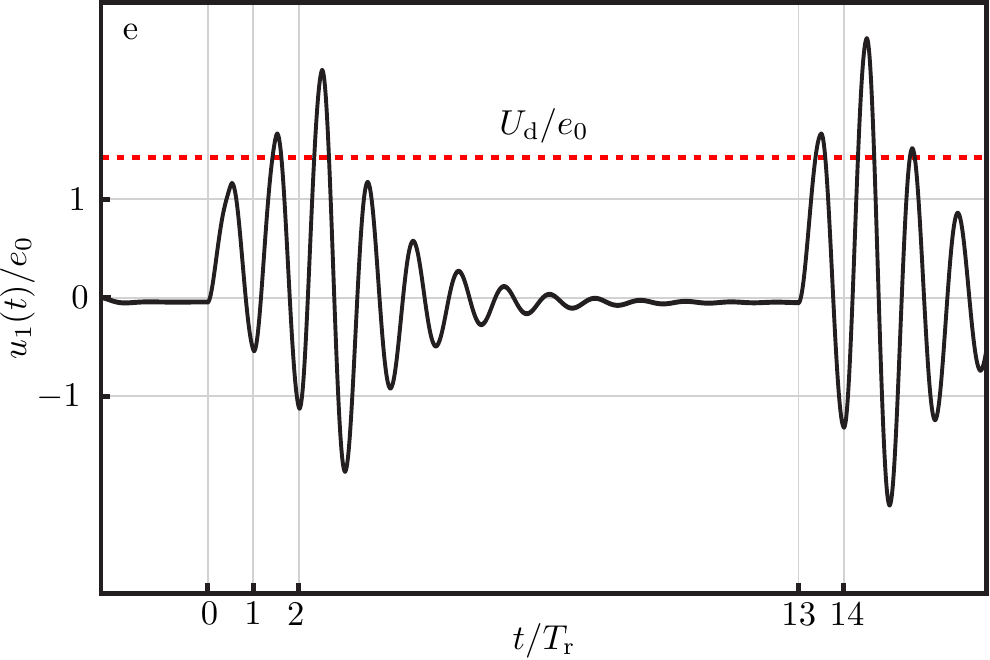}\\[0.45cm]%
		\includegraphics[width=0.4\textwidth]{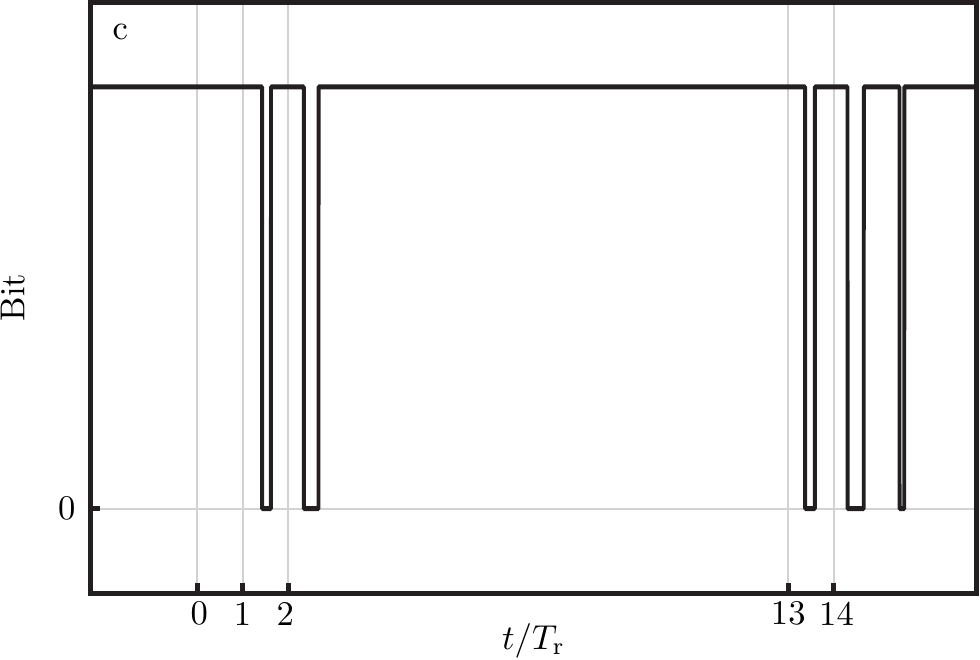}\qquad%
		\includegraphics[width=0.4\textwidth]{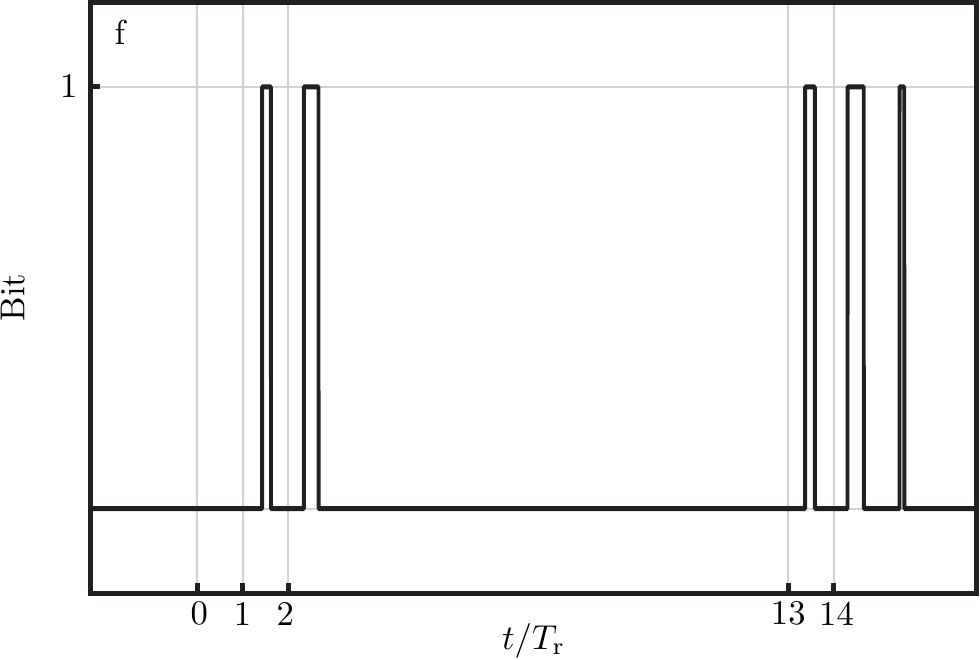}%
	\end{center}
	\caption{Simulation results of the bit anticipation circuit, where the parameters $e_0 = 2.1~\text{V}$, $U_{\mathrm{d}} = 3.8~\text{V}$, $R = 100~\Omega$, $C = 150~\text{nF}$, $L = 100~\text{mH}$, $\beta=10^{16}$, $U_{\text{on}}=1.4~\text{V}$, $U_{\text{off}} = -0.1~\text{V}$, $M_{\text{on}} = 1~\text{k}\Omega$ and $M_{\text{off}} = 0.5~\text{G}\Omega$ have been used. Negative~(a) and positive~(d) voltage pulses were used for anticipating a bit~$0$ and $1$, respectively. Depending on the excitation signal there are oscillating output voltages $u_0(t)~\mathrm{for~bit~0}$ and $u_1(t)~\mathrm{for~bit~1}$ (b, e). A bit~$0$ is anticipated for negative excitation pulses (c), whereas a bit~$1$ is anticipated for positive excitation pulses (f).}
	\label{fig:simulationResults}
\end{figure*}

The simulation results are shown in Fig.~\ref{fig:simulationResults}: After a certain sequence of learning process, the electrical circuit is indeed capable of anticipating a bit value of 0 for negative pulses and a bit value of 1 for positive pulses. In order to digitize the analogue output voltage a threshold voltage of $U_{\mathrm{d}_0} = -U_\mathrm{d} = -3.8$~V and $U_{\mathrm{d}_1} = U_\mathrm{d}$ was chosen for a bit value of 0 and 1, respectively. Under the conditions that have been mentioned in the context of (\ref{eqn:pulsetrain}) the simulation result shows, that for negative pulses the amplitude of the oscillating voltage $u_0\left(t\right)$ and for positive pulses the amplitude of the oscillating voltage $u_1\left(t\right)$ grows up. After the learning process, one pulse is sufficient to yield an oscillating voltage with high amplitude at the associated output terminal. Therefore the anticipation circuit enables bit anticipation.

\subsection{Measurements}
\begin{figure*}[!hbt]
	\begin{center}
		\includegraphics[width=0.4\textwidth]{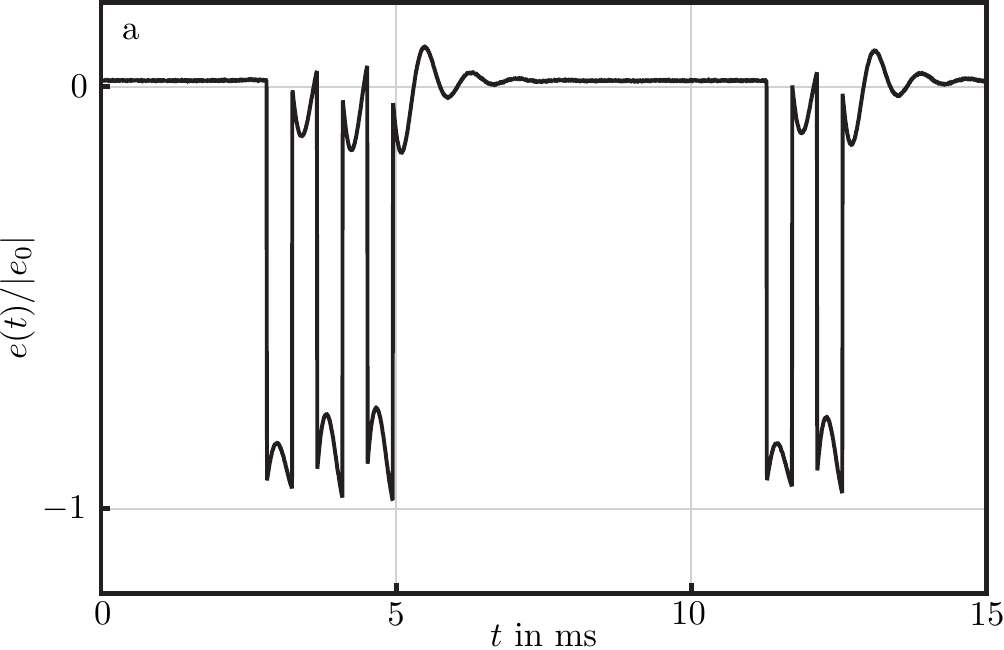}\qquad%
		\includegraphics[width=0.4\textwidth]{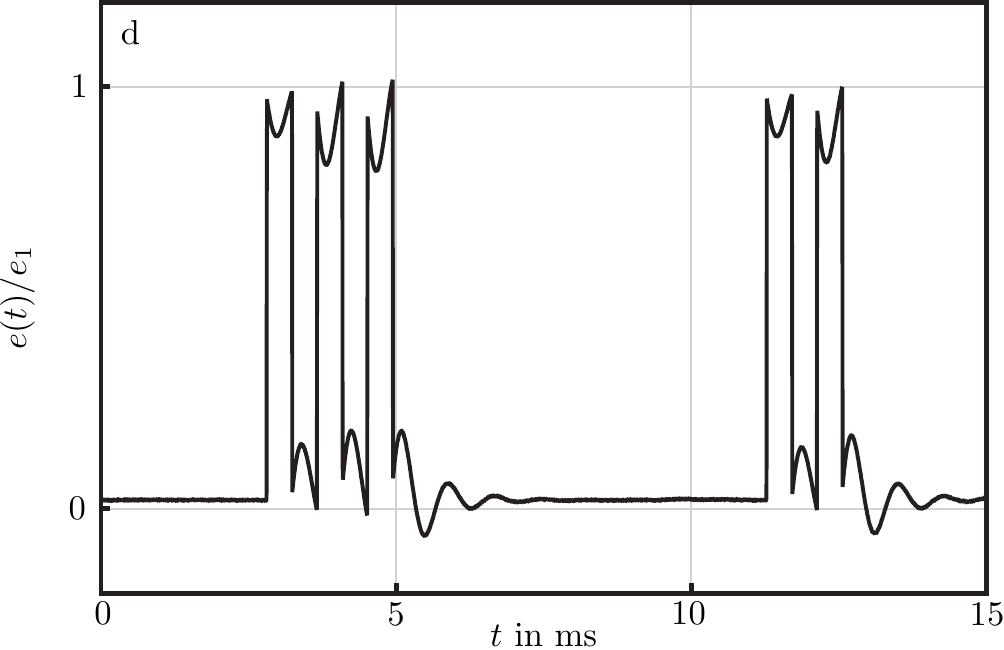}\\[0.45cm]%
		\includegraphics[width=0.4\textwidth]{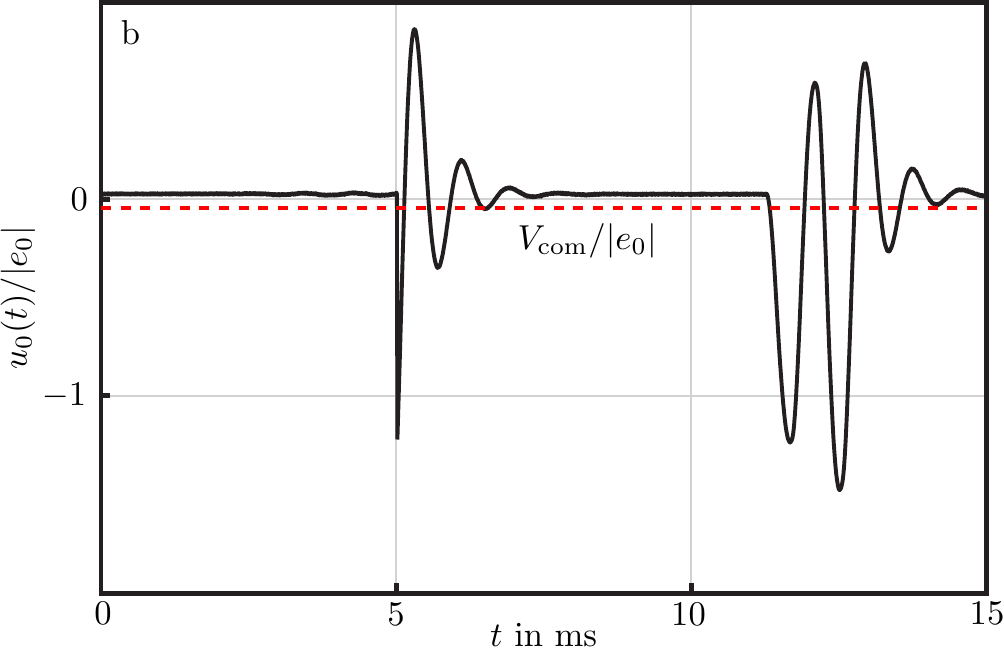}\qquad%
		\includegraphics[width=0.4\textwidth]{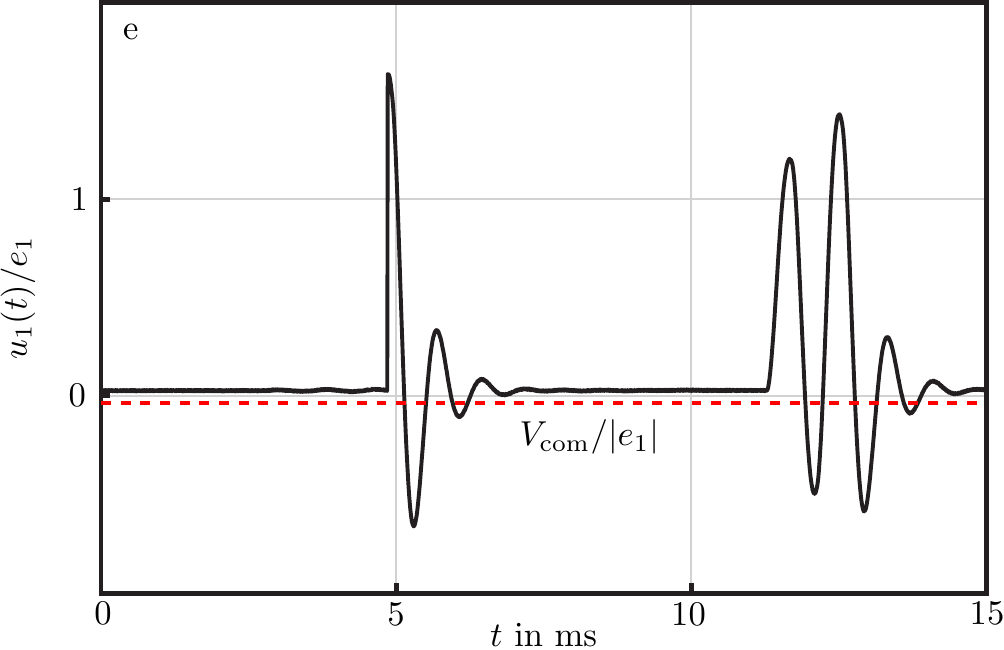}\\[0.45cm]%
		\includegraphics[width=0.4\textwidth]{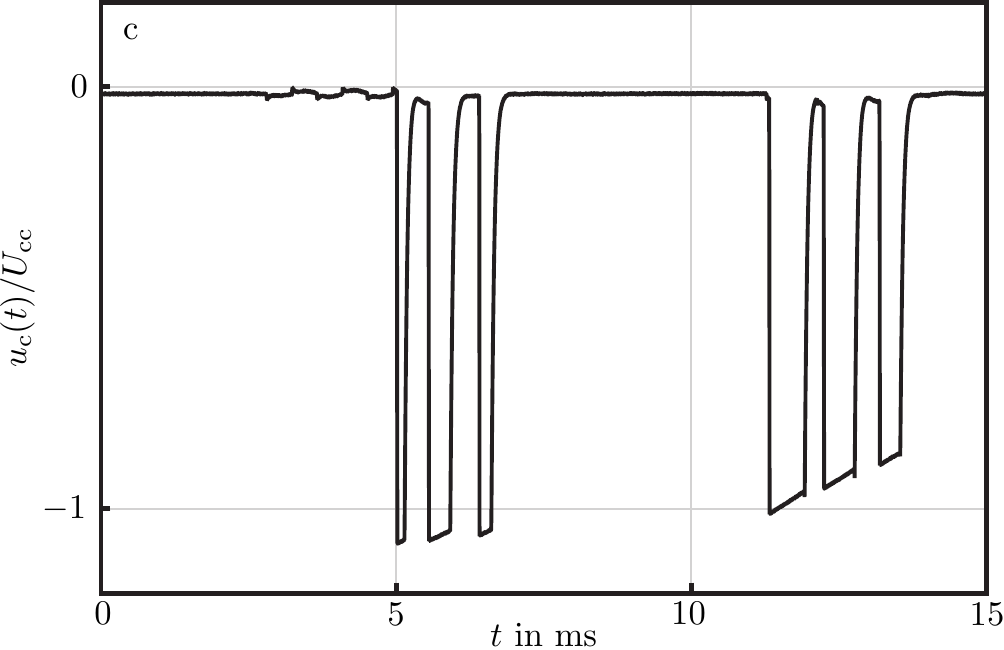}\qquad%
		\includegraphics[width=0.4\textwidth]{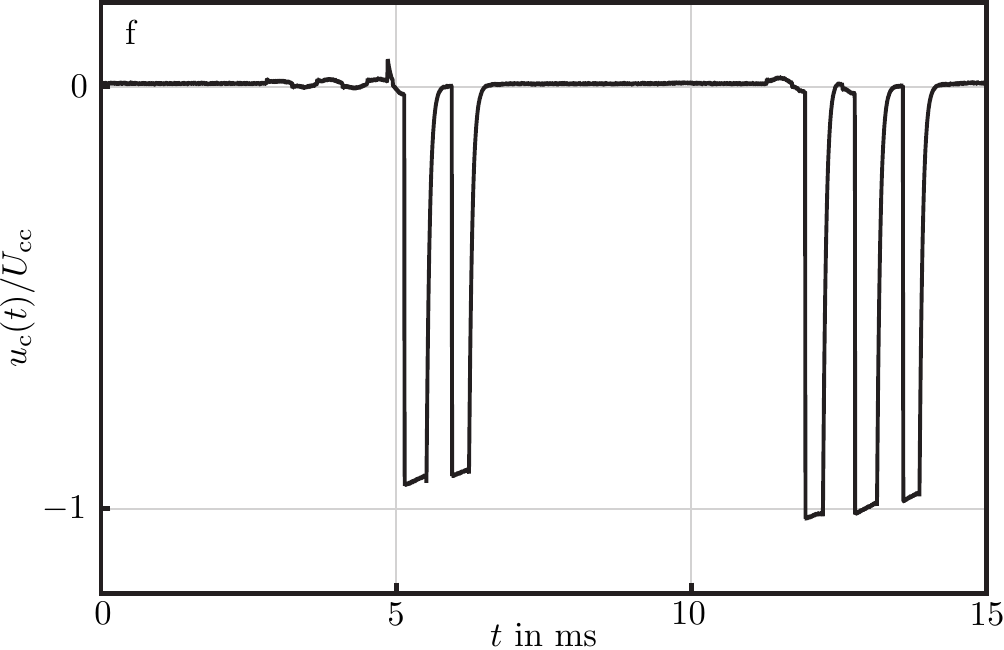}%
	\end{center}
	\caption{Measured signal sequences of the bit anticipation circuit in accordance to Fig.~\ref{fig:simulationResults}: Input voltage pulse sequence for the anticipation of a bit~$0$ (a), the corresponding oscillating output signal (b) and the anticipated logical states (d). In (d,e,f) the signals for the anticipation of a bit~$1$ are shown. Here, $U_\mathrm{cc}$ denotes the comparator output voltage amplitude and $V_\mathrm{com}$ is the comparator threshold voltage ((b, e) red-dashed).}
	\label{fig:ExperimentSignals}
\end{figure*}

The measured signals of the electrical implemented bit anticipation circuit are depicted in Fig.~\ref{fig:ExperimentSignals}. The bipolar voltage trains, which were applied to the circuit to anticipate digital patterns, are shown in Fig.~\ref{fig:ExperimentSignals}(a,d). Both voltage trains contain three learning voltage pulses in series at the resonance frequency of the circuit followed by a set of two separated voltage pulses. The second set of voltage pulses represents an incomplete bit pattern with only two pulses, which were used to demonstrate the pattern completion ability of the anticipation circuit.

At each single voltage input pulse the voltage across the memristive device increases, where at a certain point the voltage across the memristive device is large enough that the corresponding device changes its resistance state from $M_\mathrm{off}$ to  $M_\mathrm{on}$. For the electronic circuit the amplitudes $e_0$ and $e_1$ of the bipolar input signals have been chosen to $e_0=-1.9\,\text{V}$ and $e_1=2.1\,\text{V}$, respectively. Those amplitude values ensure that the voltage oscillation at the third input voltage pulse is large enough to overcome the set voltage $U_{\text{on}}$ of the memristive device. Therewith, the particular memristive device switches its resistance state and the given bit pattern is anticipated. The voltages $u_0$ and $u_1$ across the resistances $R_{2,1}$ and $R_{2,2}$ are shown in Fig.~\ref{fig:ExperimentSignals}(b,e). In the beginning the memristances $M_1$ and $M_2$ are much higher as the series resistances $R_{2,1}$ and $R_{2,2}$. Consequently, most of the input voltage drops across the particular memristive device, such that $u_0$ and $u_1$ remains nearly unaffected. The voltage $u_0$ or $u_1$ alters significantly if the corresponding bit is anticipated. After the learning intervals, the incomplete pattern of only two input pulses results in an oscillation of $u_0$ or $u_1$, in which a negative input pulse yields an oscillating voltage $u_0$, while a positive input pulse results in an oscillation of $u_1$, cf.~Fig.~\ref{fig:ExperimentSignals}(b,e)). Finally, these signals are digitized using the comparator electronics, see Fig.~\ref{fig:ExperimentSignals}(c,f). We like to remark that the threshold voltage in the experimental realization have been chosen with respect to the voltage dividers $R_{2,1}$ and $M_1$ or $R_{2,2}$ and $M_2$. In contrast to the simulation model, the voltage used for the comparator electronics differs therefore only from zero, if the respective memristive device is in the low ohmic state. Hence, a smaller voltage (labeld by $V_\mathrm{com}/e_0$ in Fig.~\ref{fig:ExperimentSignals}) has been used compared to the simulation, cf. Fig.~\ref{fig:simulationResults}. To summarize, the circuit enables a kind of pattern completion, where after the learning phase an incomplete pattern, which consists of only two voltages pulses, results in an output pattern consisting of three individual voltage pulses.


\subsection*{Demonstration of a software bit-array anticipator}
So far the bit anticipator is understood as a basic element and has been discussed in an abstract way. Now, we will focus on the anticipation of information, where any information can be represented by a set of bits, which in turn can be anticipated by an array of bit anticipators. For the sake of clarity, we used a grayscale image of digits with $5\times9$ pixels as input information. Black and white pixels are associated with the bits $0$ and $1$, respectively. Gray values in between have to be quantized in order to decide if this value belongs to a white or black pixel. In order to visualize an analog information change under certain conditions, we refrain from this quantization.

At first, we train the anticipation array by the images in accordance to the excitation used in Fig.~\ref{fig:simulationResults}. As expected, after the first three input training images the output image fades out over time, but the image is successfully anticipated after the last two excitations. Hence, anticipation can be regarded as to regenerate information.

\begin{figure*}[!h]
	\begin{center}
		\includegraphics[width=0.8\textwidth]{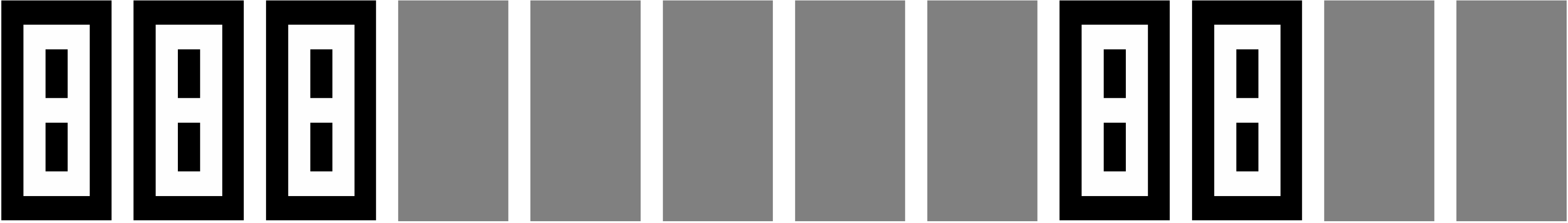}\\%
		\includegraphics[width=0.8\textwidth]{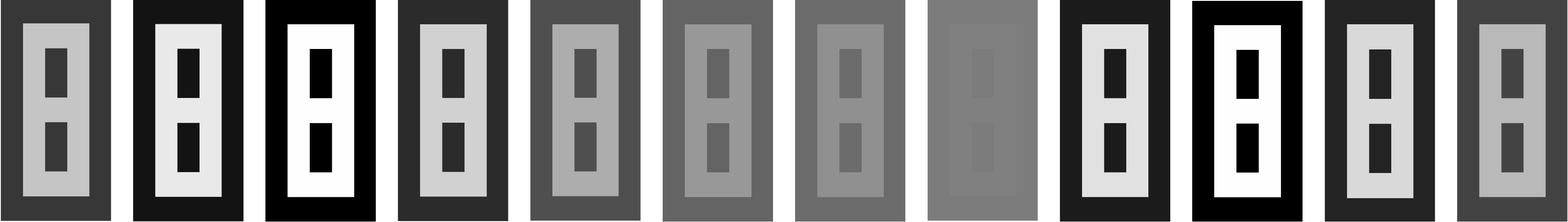}\\[0.5cm]%
	\end{center}
	\caption{Anticipation of images: Input sequence (top) and output sequence (bottom).}
	\label{fig:deterministic}
\end{figure*}

\begin{figure*}[!hbt]
	\begin{center}
		\includegraphics[width=0.8\textwidth]{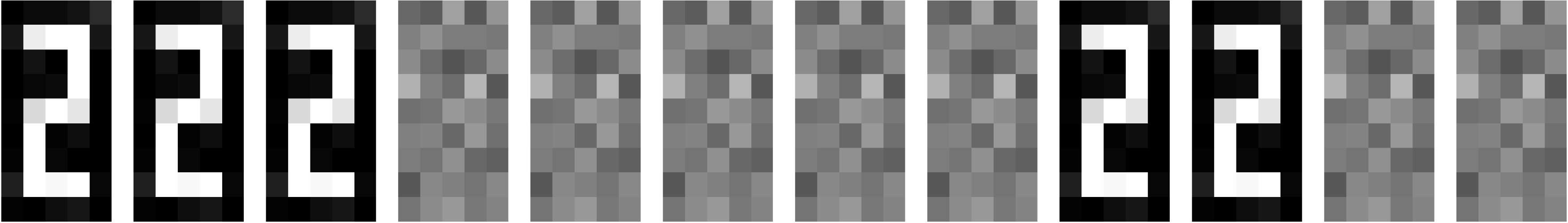}\\%
		\includegraphics[width=0.8\textwidth]{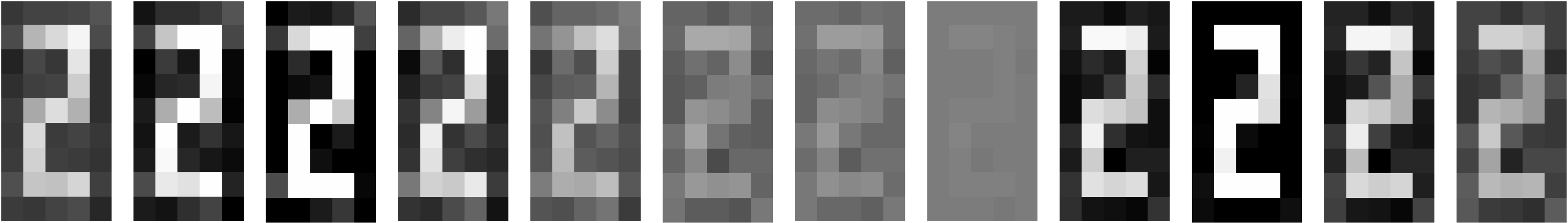}\\[0.5cm]%
	\end{center}
	\caption{Anticipation of noisy images: Input sequence (top) and output sequence (bottom).}
	\label{fig:stochastic}
\end{figure*}

Secondly, we examine the anticipation under noisy conditions. From Fig.~\ref{fig:stochastic} it can be taken that this anticipatory system inherently reduces the noise and the anticipation is still working. Of course, this depends on the noise level. In the case of a higher noise level, further simulations have shown that this can be compensated by increasing the number of training images.

\begin{figure*}[!hbt]
	\begin{center}
		\includegraphics[width=0.8\textwidth]{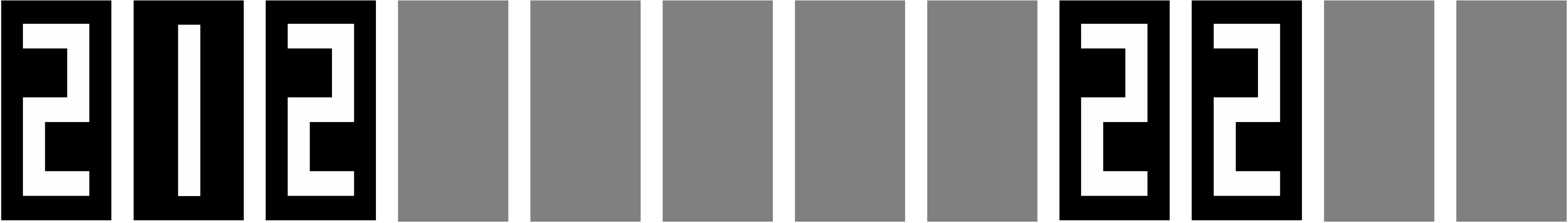}\\%
		\includegraphics[width=0.8\textwidth]{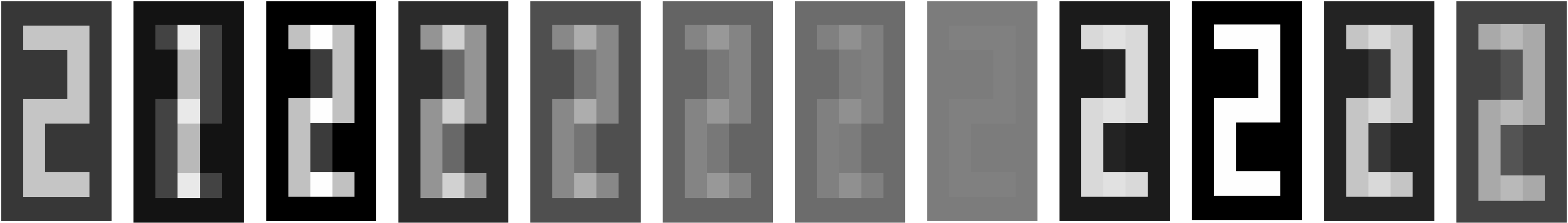}\\[0.5cm]%
	\end{center}
	\caption{Anticipation of images disturbed by a failure image: Input sequence (top) and output sequence (bottom).}
	\label{fig:intermediate}
\end{figure*}

At last, we inspect the anticipation, when the input sequence is disturbed by an intermediate failure image, see Fig.~\ref{fig:intermediate}. From the output sequence, it can be verified that the failure image is treated like in the noisy case before. When it comes to the second excitation with the two correct images the influence of the failure image can be neglected, i.\,e. this anticipatory system has filtered out the faulty information.

\section{Conclusions}
Anticipation is a general concept in nature and even observed in unicellular creature such as amoebas. In this work, a bio-inspired memristive circuit is presented which enables information anticipation. It is based on a memristive resonator circuit, which mimics the anticipation capability of an amoeba but is limited to anticipate unipolar pulses. Moreover, this circuit has been extended to anticipate bipolar pulses representing an information bit with the values $0$ or $1$. This extension allows for the anticipation of a single bit, which is the basic unit of information in digital computing. An anticipation of bit vectors can be realized by replicating this circuit pursuant to the length of the bit vector. Hence, any information represented by such a bit vector -- which occurs in periodical events -- can be anticipated. In particular, this functionality could be interesting in robotics. For instance, if this bit vector contains information about location, velocity vectors, and so on, a robot could predict a future location and act according to this prediction. To sum up, our work can be considered as a path towards implementing the biological phenomenon of anticipation in technical systems by using a memristive circuitry, which is capable of pattern completion even under noisy signal conditions or false input information.

\section*{Acknowledgement}
We acknowledge support by: German Science Foundation (DFG) - Research Group FOR 2093
'Memristive Bauelemente f\"{u}r neuronale Systeme'.

\end{document}